\documentclass[aps,pre,preprint]{revtex4-1}
\usepackage{epsf}
\usepackage{latexsym}
\usepackage{epsfig}
\usepackage{amsmath}
\usepackage{amsfonts}
\usepackage{amssymb}
\usepackage{graphicx}
\usepackage{color}
\usepackage[usenames,dvipsnames]{xcolor}

\newcommand{\prs}[1]{{\left(#1\right)}}
\newcommand{\col}[1]{{\left[#1\right]}}
\newcommand{\chs}[1]{{\left\{#1\right\}}}
\newcommand{\avg}[2]{{\left<#1\right>_{#2}}}
\newcommand{\prob}[1]{{\mathcal{P}\prs{#1}}}
\newcommand{\pder}[2]{{\frac{\partial #1}{\partial #2}}}

\newcommand{\vc}[1]{{\boldsymbol #1}}
\newcommand{\atanh}{{\mbox{atanh }}}

\newcommand{\sgn}{{\mbox{sgn }}}
\newcommand{\erf}{{\mbox{erf}}}

\newcommand{\cut}[1]{{}}

\begin{document}

\title{Replication-based Inference Algorithms for Hard Computational Problems}
\author{Roberto C. Alamino, Juan P. Neirotti and David Saad}
\affiliation{Non-linearity and Complexity Research Group, Aston University, Birmingham B4 7ET, UK}

\begin{abstract}
Inference algorithms based on evolving interactions between replicated solutions are introduced and analyzed on a prototypical
NP-hard problem - the capacity of the binary Ising perceptron. The efficiency of the algorithm is examined numerically against
that of the parallel tempering algorithm, showing improved performance in terms of the results obtained, computing requirements 
and simplicity of implementation.
\end{abstract}

\maketitle

\section{Introduction}

One of the main contributions of statistical physics to application domains such as information theory and theoretical computer
science has been the introduction of established methods that facilitate the analysis of typical properties of very large systems
in the presence of disorder. For instance, in information theory applications, these large systems correspond to (mostly binary)
transmissions where the disorder is manifested through transmission noise or the manner by which the message is generated or
encoded. Established approaches in the statistical physics community such as the replica and cavity methods~\cite{mezard09}
have proved to be useful tools in describing typical properties of error-correcting codes~\cite{Saad01,Alamino07,Alamino07b},
the analysis of optimization problems such as the traveling salesman~\cite{Mezard86}, K-satisfiability~\cite{Braunstein05} and
graph coloring~\cite{vanMourik02,mulet02} to name but a few.

Another important contribution, which complements the ones mentioned above, was in the development of algorithmic tools to find
microscopic solutions in specific problem instances. One of the most celebrated inference methods, the message-passing (MP) or
belief propagation algorithm, had been developed independently in the information theory~\cite{Gallager63},
machine learning~\cite{Pearl88} and statistical physics~\cite{Mezard86} communities until the links between them have been
identified~\cite{Kabashima98,Opper01} and established~\cite{yedidia05}. Subsequently, a number of successful inference methods
have been devised using insights gained from statistical physics~\cite{mezard02b,mezard09}.

In MP algorithms, the system to be solved is mapped onto a bipartite factor graph, where on the one hand factor nodes correspond
to observed (given) information or interaction between variables; while on the other hand are variable nodes, to be estimated on
the basis of approximate marginal pseudo-posteriors. The latter are obtained by a set of consistent marginal conditional 
probabilities (messages) passed between variable and factor nodes~\cite{mezard09}. Unfortunately there are many caveats to the MP 
procedure, especially in the presence of closed loops in the factor graph, which may give rise to inconsistent messages and 
non-convergence. It can be shown that MP converges to the correct solution when the factor graph is a tree but there is no such 
guarantee for more general graphs, although MP does result in good solution in many other cases too.

There are two main general difficulties in using MP algorithms to problems represented by densely connected graphs. The first is
that the computational cost grows exponentially with the degree, making the computation impractical; while the second arises
from the existence of many short loops that result in recurrent messages and lack of convergence. These problems have been solved
in specific cases, especially in the case of real observations and continuous noise models by aggregating
messages~\cite{Kabashima03}. One of the shortcomings identified in~\cite{Kabashima03} was non-convergence when prior knowledge
on the noise process is inaccurate or unknown; which typically results in multiple solutions and conflicting messages.

While MP would be successfully applied if a weighted average over \emph{all} possible states could be carried out, it is clear
that such average is infeasible. Inspired by the state-space representation obtained using the replica
method~\cite{Mezard86,nishimori01,mezard09} whereby state vectors are organized in an ultrametric structure, two of us suggested
an MP algorithm based on averaging messages over a structured solutions space~\cite{Neirotti05}. The approach is based on using
an infinite number of copies (or \emph{real replica}, not to be confused with those employed in the replica method) of the
variables exposed to the same observations (factor nodes). The replicated variable systems facilitate a broader exploration of
solution space as long as these replica are judiciously distributed according to the solution-space structure implied by the
statistical mechanics analysis. The variable vectors inferred by these algorithms are then combined by taking either weighted or
white average to obtain the marginal pseudo-posterior of the various variables.

The approach has been successful in addressing the Code Division Multiple Access (CDMA) problem as well as the Linear Ising
Perceptron capacity problem~\cite{Neirotti07}, even in cases where prior information is absent.
It is worthwhile noting that a seed of this replication philosophy can be found in several previous algorithms such
as: (a) \emph{query by committee}~\cite{Seung92}, where the potential solutions (system replica) are used for choosing the best
most informative next example and later combines the solutions using a majority voting. (b) An analytical
approach~\cite{deDominicis80} aimed at obtaining solutions for the Sherrington-Kirkpatrick model via averages over the
Thouless-Anderson-Palmer equations. (c) A study of the p-spin model meta-stable states by considering averages over a small
number of real replica~\cite{Kurchan93}. (d) The \emph{Parallel Tempering} (PT) algorithm, also known as replica exchange MCMC
sampling, which relies on many replica searching the space at different temperatures~\cite{Swendsen86,Marinari1992}; the latter,
due to its good performance and relation to the approach we advocate, will be explained in more detail later on and will be used
for comparison with the method develop here. It is interesting to note that approaches  based on averaging multiple interacting
solutions have also been successfully tried in neighboring disciplines, for example for decoding in the context of
error-correcting codes~\cite{Shrinivas2011}.

While this approach has been successful in addressing inference problems in the case of real observations and continuous noise
models, it is less clear how it could be extended to accommodate more general cases. In this work we will present an alternative
method for carrying out averages over the replicated solutions, which can be applied to more general cases. Generally, like most
MP-based algorithms, the approach is based on solutions being calculated iteratively using a pair of coupled self-consistent
equations. We will study the properties of the new algorithm, its advantages and limitations, on an exemplar problem of the
Binary Ising Perceptron (BIP)~\cite{Krauth89,Engel01} that has been used as a benchmark also in other works on advanced inference
methods~\cite{Braunstein06}.

One obvious obstacle in most MP algorithms is that the iterative dynamics can be trapped in suboptimal minima;
in addition, the algorithm itself can either create spurious suboptimal minima in the already complex solution space
or change the height of the energy barriers between the existing ones. We will show that our replica-based MP algorithm fails
under naive averaging of the replica for the BIP capacity problem, explain analytically why it happens and show that in the limit
of a large number of replica, averages flow to the clipped Hebb algorithm~\cite{Engel01}. We will then propose an alternative
approach and show how replication can indeed improve performance if carried out appropriately.

In section~\ref{section:BIP} we will explain the exemplar problem to be used in this study; we will then review the non-replicated
MP solution to the BIP capacity problem under the approximation for densely connected systems in section~\ref{section:MP} and
provide an analytical solution to the naively replicated MP algorithm, showing here its equivalence with the clipped Hebb rule.
Section~\ref{section:OMP} will point to the main reason for the failure of the naive replica-averaging approach and argue that an
online version of the MP algorithm, which is derived and presented, can solve it. By replicating the new online MP (OnMP)
algorithm and using the extra degrees of freedom that it provides, we show how it outperforms the non-replicated MP algorithm,
termed offline MP (OffMP) algorithm. Section~\ref{section:CP} compares the replicated OnMP (rOnMP) with
a benchmark parallel algorithm, namely the PT algorithm. Finally, conclusions and future directions are discussed in
section~\ref{section:Conclusions}.

\section{Exemplar Problem - the Binary Ising Perceptron}
\label{section:BIP}

To extend the replica-based inference method~\cite{Neirotti07} we would like to use an exemplar problem that is particularly
difficult, not only in the worst-case scenario but also typically, where both observations and noise model are not real-valued.
In addition, we would like to examine a case where exact results have been obtained by the replica theory; this provides helpful
insight in devising the corresponding algorithm by suggesting a possible structure for the solution space as well as an analytical
tool to assess the efficacy of the algorithm.

One prototypical NP-complete problem~\cite{Pitt88} that was shown to be computationally hard even in the typical case, which was
solved exactly using the replica method, is the capacity of the Binary Ising Perceptron (BIP)~\cite{Krauth89}. This is due to the
complex structure of its solution space studied in~\cite{Obuchi09}, showing a non-trivial topology even in the replica symmetric
(RS) phase.

The BIP~\cite{Engel01} represents a process whereby $K$-dimensional binary input vectors $\vc{s}_\mu\in\chs{\pm1}^K$ are
received, where the input vector index $\mu=1,...,N$, represents each of the $N$ example vectors. The corresponding outputs for
each one of them is determined by the binary classification
\begin{equation}
  y_\mu = \sgn\prs{\frac1{\sqrt{K}}\sum_{k=1}^K s_{\mu k} b_k},
\end{equation}
where $\vc{b}=(b_1,...,b_K)\in\chs{\pm1}^K$ is called the unknown binary variables (also referred to as the perceptron's variable
vector); the prefactor $\sqrt{K}$ is for scaling purposes, so that the argument of the sign function remains order
$O(1)$ as $K\rightarrow\infty$.

The capacity problem for a BIP is a storage problem, although it can alternatively be seen as a compression
task~\cite{Hosaka2002}. In the simplest version of the problem, a dataset $D=\chs{(\vc{s}_\mu,y_\mu)}_{\mu=1}^N$ consisting of
$N$ pairs of inputs and outputs (also called \emph{examples}) is randomly generated and a perceptron with an appropriate variable
vector $\vc{b}$ should be found, such that when presented with an input pattern $\vc{s}_\mu$, it reproduces the corresponding
output $y_\mu$. That is the equivalent of compressing the information contained in the set of classifications $\chs{y_\mu}$,
comprising $N$ bits, into a vector $\vc{b}$ with only $K$ bits. One is usually interested in the typical case, which is
calculated by averaging over all possible datasets $D$ drawn at random from a certain probability distribution.

Typical performances are algorithm dependent and are measured by counting the fraction of correctly stored patterns as a function
of the number of examples in the dataset. One very convenient measure used in statistical physics calculations is the average
value of the energy function
\begin{equation}
  E (\hat{\vc{b}})= 1-\prod_{\mu=1}^N \Theta\prs{y_\mu \frac1{\sqrt{K}} \sum_{k=1}^K s_{\mu k} \hat{b}_k},
  \label{equation:Energy}
\end{equation}
with $\Theta(\cdot)$ being the Heaviside step function and $\hat{\vc{b}}$ the inferred variable vector. This measure gives 0 if
all examples are correctly learned and 1 otherwise, i.e., it is an indicator that all the patterns were perfectly
memorized. The maximum value of $\alpha=N/K$ for which this cost function is 1 (when averaged over all possible datasets) is
the \emph{achieved capacity} of the algorithm and a measure of its overall performance.

Although the achieved capacity varies between algorithms, there is an absolute upper bound, the critical capacity $\alpha_c$,
above which no algorithm can memorize the whole set of examples in the typical case (although it might be possible for specific
instances); this reflects the information content limit of the perceptron itself.

The critical capacity was calculated by Krauth and M\'{e}zard using the one-replica symmetry breaking (1RSB)
ansatz~\cite{Krauth89} with the result of $\alpha_c\approx0.83$. Taking into consideration that the problem is computationally
hard, the challenge then becomes to find an algorithm which infers appropriate $\vc{b}$ values in typical specific instances of 
$D$ as close as possible to $\alpha_c$, where the corresponding computational complexity scales polynomially with the system size.

\section{Naive Message Passing}
\label{section:MP}

The inference problem we aim to address is finding the most appropriate value of the variable vector $\vc{b}$
capable of reproducing the classifications given the examples dataset $D$.
Firstly, one needs to determine a quality measure that quantifies the appropriateness of a solution.
The most commonly used error measure in similar estimation problems is the expected error per variable, or bit-error-rate in the
information theory literature, the minimization of which leads to a solution based on the Marginal Posterior Maximiser (MPM) 
estimator given by
\begin{equation}
  \hat{b}_k = \mbox{argmax}_{b_k\in\chs{\pm1}} \, \sum_{b_{l\neq k}} \prob{\vc{b}|D} = \sgn \avg{b_k}{\prob{\vc{b}|D}},
\end{equation}
which means that one estimates $\vc{b}$ bitwise, such that each component $\hat{b}_k$ corresponds to the variable value that
maximizes the marginal distribution per variable given the dataset $D$. The MP equations allow one to carry
out an approximate Bayesian inference procedure to find this estimator.

It is important to remember that there might not exist a variable vector capable of reproducing the whole dataset. In
this case, the dataset is \emph{unrealizable} by the BIP, although one can still identify the most probable candidate.
In the BIP capacity problem, unrealizable datasets exist since they are generated randomly, not by a teacher perceptron
as is the case in some generalization problems.
Each variable in the set $D$ is drawn from an independent distribution and therefore one can write the posterior
distribution of the variable vector as
\begin{equation}
 \prob{\vc{b}|D} = \prob{\vc{b}|\chs{y_\mu},\chs{\vc{s}_\mu}} \propto \prob{\chs{y_\mu}|\vc{b},\chs{\vc{s}_\mu}}\prob{\vc{b}},
\end{equation}
where $\prob{\chs{y_\mu}|\vc{b},\chs{\vc{s}_\mu}}$ factorizes as the examples are sampled identically and independently
\begin{equation}
  \prob{\chs{y_\mu}|\vc{b},\chs{\vc{s}_\mu}}=\prod_{\mu=1}^N \prob{y_\mu|\vc{b},\vc{s}_\mu}.
\end{equation}
From the Bayesian point of view, $\prob{\vc{b}}$ is interpreted as the (factorized) prior distribution of possible variable
vectors. As there is no noise involved in the capacity problem, the likelihood factor is simply given by
\begin{equation}
  \prob{y_\mu|\vc{b},\vc{s}_\mu} = \frac12+\frac{y_\mu}2 \sgn\xi_\mu,
\end{equation}
defining
\begin{equation}
  \xi_\mu = \frac1{\sqrt{K}}\sum_{k=1}^K s_{\mu k} b_k.
\end{equation}
As for each instance the dataset is fixed, we will omit in the following expressions the explicit reference to the input
vectors $\vc{s}_\mu$ in the posterior distribution for brevity.

The resulting MP equations are self-consistent coupled equations of marginal conditional probabilities
which are iterated until convergence (or up to a cutoff number of iterations). These equations are obtained by applying Bayes
theorem to each one of the so-called $Q$-messages and $R$-messages
\begin{align}
  Q^{t+1}_{\mu k} \prs{b_k} &= \mathcal{P}^{t+1} \prs{b_k|\chs{y_{\nu\neq\mu}}}
                               \propto \prob{b_k} \prod_{\nu\neq\mu} \mathcal{P}^{t+1} \prs{y_\nu|b_k,\chs{y_{\sigma\neq\nu}}}\\
  R^{t+1}_{\mu k} \prs{b_k} &= \mathcal{P}^{t+1} \prs{y_\mu|b_k,\chs{y_{\nu\neq\mu}}}
                               =\sum_{\chs{b_{l\neq k}}} \prob{y_\mu|\vc{b}}\prod_{l\neq k} \mathcal{P}^t
                               \prs{b_l|\chs{y_{\nu\neq\mu}}},
\end{align}
where $\prob{b_k}$ is the prior distribution over the $k$-th entry of the variable vector and $t$ stands for the current
iteration step.

As $b_k\in\chs{\pm1}$, one can write
\begin{equation}
  Q^t \prs{b_k} =       \frac{1+m^t_{\mu k}b_k}2 \qquad \text{and} \qquad
  R^t \prs{b_k} \propto \frac{1+\hat{m}^{t-1}_{\mu k}b_k}2.
\end{equation}

The variables $m_{\mu k}$ may be interpreted as magnetization related to the cavity field in analogy to
spin lattices in magnetic fields. The interpretation of the $\hat{m}_{\mu k}$ variables is less intuitive. Substituting the
$R$-messages into the $Q$-messages and summing over the two possible values of $b_k$ we finally reproduce the MP equations in
their well-known form
\begin{align}
  \label{equation:orsp_1}
  \hat{m}^t_{\mu k} &= \frac{\sum_{b_k} b_k \mathcal{P}^{t+1}\prs{y_\mu|b_k,\chs{y_{\nu\neq\mu}}}}
                       {\sum_{b_k}\mathcal{P}^{t+1}\prs{y_\mu|b_k,\chs{y_{\nu\neq\mu}}}},\\
  \label{equation:orsp_2}
  m^t_{\mu k}       &= \tanh\col{\sum_{\nu\neq\mu} \atanh \hat{m}^t_{\nu k}}
                       \approx \tanh\prs{\sum_{\nu\neq\mu} \hat{m}^t_{\nu k}},
\end{align}
the approximation in the last equation is possible since $\hat{m}_{\mu k} \sim O(1/\sqrt{K})$ as we will see later.

Once convergence is attained, the value for the variable vector can be estimated by
\begin{align}
  \hat{b}_k &= \sgn m_k,\\
  \label{equation:magk}
  m_k &= \tanh\prs{\sum_\nu \hat{m}^t_{\nu k}},
\end{align}
or
\begin{equation}
  \hat{b}_k = \sgn \prs{\sum_\nu \hat{m}^t_{\nu k}}.
\end{equation}

As mentioned in section~\ref{section:BIP}, the factor graph representing the BIP is densely connected, but an expansion for large
$K$ suggested by Kabashima~\cite{Kabashima03} helps to simplify the equations away from criticality. However, for the BIP this
expansion requires extra care due to the discontinuity of the sign function. To address this problem, we developed a different
approach to carry out this expansion which can be generalized to accommodate other types of perceptrons with minimal
modifications; it can be applied to either continuous or discontinuous activation functions, with or without noise.
Equation~(\ref{equation:orsp_2}) for $m_{\mu k}$ is not modified, but $\hat{m}_{\mu k}$ is expanded in powers of $1/\sqrt{K}$,
giving rise to a different expression (detailed derivation is provided in appendix \ref{appendix:MPE}):
\begin{equation}
  \label{equation:MPFE}
  \hat{m}_{\mu k} =\frac{2 s_{\mu k} y_\mu }{\sqrt{K}}\frac{\mathcal{N}_{\mu k}}{1+\erf\prs{y_\mu  u_{\mu k}/\sqrt{2\sigma^2_{\mu k}}}},
\end{equation}
where
\begin{align}
  \mathcal{N}_{\mu k} &= \frac1{\sqrt{2\pi\sigma^2_{\mu k}}} \exp\prs{-\frac{u_{\mu k}^2}{2\sigma^2_{\mu k}}},\\
  \sigma^2_{\mu k}    &= \frac1K \sum_{l\neq k} (1-m_{\mu l}^2),\\
  u_{\mu k}           &= \frac1{\sqrt{K}} \sum_{l\neq k} m_{\mu l} s_{\mu l}.
\end{align}

However, this version of the algorithm is unable to memorize large numbers of examples. Simulation results show that,
even for small system sizes ($K\sim10$) it cannot memorize more than a single pattern on average. This is
a consequence of the fact that the dynamical map defined by the MP equations becomes trapped in the many suboptimal minima
of the energy landscape.

In principle, one should be able to correct this by replicating the system and distributing the $n$ replica randomly in
solution space, let each one carry out the inference task independently and compare their final fixed points. An idea along these
lines, with a small number of real replica searching the space in parallel, was tested with some success in~\cite{Kurchan93};
where the replica helped change the landscape to facilitate jumps over barriers between metastable states. However, the
corresponding algorithm was not very efficient computationally. Also the replicated version of MP we tested failed and the
observed performance coincided with that of the non-replicated version.

To understand the reasons for the failure of the naively replicated algorithm, we solved the replicated version of the algorithm
analytically. We consider the case where a simple white average of the $n$ replica is used for inferring the variable vector value
\begin{equation}
  \hat{b}_k = \sgn\prs{\frac1n \sum_{a=1}^n b_k^a}~.
\end{equation}

We can then evaluate the MP equations using a saddle point method when $n,N,K\rightarrow\infty$. The detailed calculation is
given in appendix~\ref{appendix:Hebb} giving rise to a surprisingly simple final result
\begin{equation}
  \label{equation:Hebb}
  \hat{b}_k = \sgn\prs{\sum_{\mu=1}^N y_\mu s_{\mu k}}.
\end{equation}

Simplicity is not the only surprising aspect of this result. Those familiar with past research in machine learning will readily 
recognize this equation as the clipped version of the Hebb learning rule~\cite{Koehler90}. Unfortunately, this is not good news 
as the maximum attainable capacity by this algorithm has been already calculated analytically
to be $\frac{N_H}{K} \equiv\alpha_H =  \approx0.11$ \cite{Sompolinsky86,vanHemmen87}. Worse yet, the achieved capacity of the
clipped-Hebb rule quickly deteriorates as $K$ increases, converging to zero asymptotically.

The flow of the replicated algorithm towards the clipped Hebb rule points out some other weaknesses of the MP algorithm.
It is not difficult to appreciate that MP results in a clipped rule as the final estimate of the variable vector is
obtained by clipping the fixed point of the magnetization; this implies that it suffers from all pathologies present in
clipped rules such
as suboptimal solutions.

Another characteristic that is highlighted by this result is the fact that, like the Hebb rule, the MP approach is an offline
(batch) learning algorithm in the sense that it does not depend on the order of presentation of the examples. This is true both
for the non-replicated and replicated algorithms. This suggests that one could introduce an extra source of stochasticity by
devising an online version of the MP, which could allow for the algorithm to overcome the energy barriers that trap it in
local minima. Different orders of examples correspond to different paths in solution space which, combined, could
potentially explore it much more efficiently. The examples order is an extra degree of freedom that cannot be exploited in
offline algorithms. In the following section we show that by pursuing this idea, we find a replicated version of MP which
does not
only perform better than the offline one (OffMP), but also offers many additional advantages.

\section{Online Message Passing}
\label{section:OMP}

The results of the previous section indicate that replication of the OffMP algorithm does not offer any
significant improvement in performance in the BIP capacity problem. The online version of the
MP algorithm introduced here allows one to exploit the order of presentation of examples as a mechanism to avoid
algorithmic trapping in local minima. This algorithm will then be used in its replicated version
with a polynomial number of replica $n$ with respect to the number of examples $N$.

In order to develop an online version of the MP algorithm we rely on a large $K$ expansion. When $K\rightarrow\infty$, one can 
derive the equations for the magnetization (mean values) of the inferred variable vector,
equation~(\ref{equation:magk}), as
\begin{equation}
\label{eq:m_online}
  \begin{split}
    m_{k} &=        \tanh \prs{\sum_{\nu} \hat{m}_{\nu k}}\\
          &=        \tanh \prs{\sum_{\nu\neq\mu} \hat{m}_{\nu k}+\hat{m}_{\mu k}}\\
          &\approx  \tanh \prs{\sum_{\nu\neq\mu} \hat{m}_{\nu k}} + \hat{m}_{\mu k}\col{1-\tanh^2 \prs{\sum_{\nu\neq\mu} 
                    \hat{m}_{\nu k}} }\\
          &=        m_{\mu k} + \col{1-\prs{m_{\mu k}}^2}\hat{m}_{\mu k}.
  \end{split}
\end{equation}

Equation~(\ref{eq:m_online}) singles out the $\mu$-th example similarly to the OffMP derivation. However, in the online 
interpretation it is considered a \emph{new example}, being presented sequentially after all previous $\mu-1$ examples 
have been learned. Then, $m_k$ can be interpreted as the updated magnetization, while $m_{\mu k}$ is the magnetization 
linked to the cavity field induced by the previous examples, before example $\mu$ is included. In the bipartite interpretation of 
the model this is akin to the introduction of new a factor node, exploiting conditional probabilities calculated with respect to 
the previous $\mu-1$ examples. To make this interpretation more explicit, we add a time label to the obtained equation by 
changing $m_k$ to
$m_k(t)$, $m_{\mu k}$ to $m_k(t-1)$ and considering the $\mu$-th example as the example being presented at time $t$.
The online MP algorithm can finally be written as
\begin{equation}
  m_k(t) = m_k(t-1) + \frac{s_{tk}y_t}{\sqrt{K}} F_k(t),
\end{equation}
with the so-called \emph{modulation function} given by
\begin{equation}
  F_k(t) = 2 \col{1-m^2_k(t-1)} \frac{\mathcal{N}_{t k}}{1+\erf\prs{y_t  u_{t k}/\sqrt{2\sigma^2_{t k}}}},
\end{equation}
where
\begin{align}
  \sigma^2_{t k} &= \frac1K \sum_{l\neq k} \col{1-m_l^2(t-1)},\\
  u_{t k}        &= \frac1{\sqrt{K}} \sum_{l\neq k} s_{t l} m_l(t-1).
\end{align}

\begin{figure}
  \center
  \includegraphics[width=12cm]{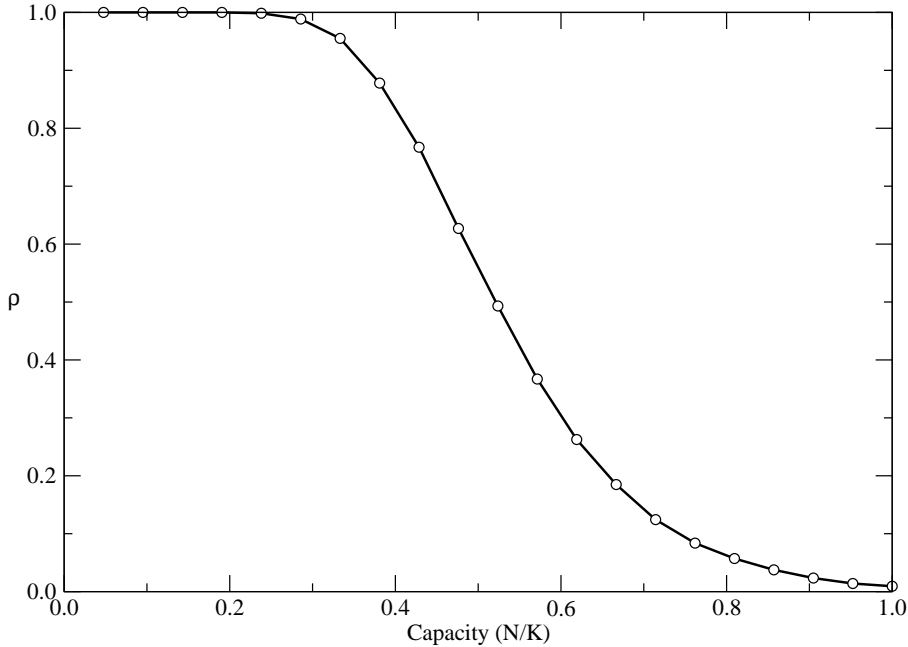}
  \caption{Non-replicated online version of the MP algorithm. While the offline MP cannot learn perfectly more than one single
           example, we see that the OnMP can, already without replication, memorize perfectly a larger number of examples. 
           The vertical
           axis $\rho$ is the average value of the indicator function that gives 1 if all patterns are memorized and zero 
           otherwise. The
           horizontal axis is the capacity $\alpha=N/K$.}
  \label{figure:OnxOff}
\end{figure}

The performance of the OnMP algorithm without replication is shown in fig. \ref{figure:OnxOff} for $K=21$ averaged over 200 
different
sets of examples. The vertical axis shows
$\rho=1-\avg{E}{}$, the average value of the function that indicates perfect learning. However, because $E\in\chs{0,1}$ we have to
estimate the variance by repeating the average several times and calculating an average over averages. For the graph of
fig. \ref{figure:OnxOff}, this was done 200 times for each particular dataset and the corresponding error bars are
smaller than the size of the symbols. We see that, contrary to the OffMP, the online version is now able to memorize perfectly a 
larger number of examples on average, with a larger achieved capacity. The slow decay to zero is to be interpreted as a finite 
size effect, which is however difficult to since increasing the system size $K$ leads to a deterioration of performance instead 
of a sharper transition.

Let us now replicate this algorithm. For $N$ examples, there are $N!$ possible orders of presentation, but we will choose only
a number $n$ of these sequences, with $n$ being of polynomial order in $N$. We will see that this is enough to improve
considerably the performance of the algorithm. We compare two versions of the replicated algorithm with white and weighted average 
over replica. Both versions work by exposing the $n$ replica independently to different orders of examples. To minimize
residual effects, we allow a relearning procedure with $L\sim 10$ relearning cycles while keeping the same order of presentation.
As the MP algorithm relies on clipping, it shows a poorer performance when the number of examples is small,
especially for the even cases where parity effects are amplified. This effect however disappears as $N$ grows larger.

The difference between the white and weighted algorithms lies in how the final estimate for the variable vector is calculated.
Respectively, we have
\begin{align}
  \hat{b}_k^{white}    &= \sgn\prs{\frac1n \sum_{a=1}^n b_k^a}\\
  \hat{b}_k^{weighted} &= \sgn\prs{\sum_{a=1}^n w^a b_k^a},
\end{align}
where
\begin{equation}
  w^a \propto e^{-\beta E(\vc{b}^a)},
\end{equation}
and the energy of each replica is calculated as in equation~(\ref{equation:Energy}). The parameter $\beta$ works as an inverse
temperature and is given a high value in order to select lower energy states. Clearly, when $\beta=0$, the white and 
weighted averages are the same.

We compared the performance of the two versions of the rOnMP against the non-replicated
one. Both \cut{of them} perform much better than the non-replicated algorithm.
The difference between weighted and white averages in related problems had already been studied in relation to
the TAP equations via the replica approach yielding similar results~\cite{deDominicis80}; this indicates that similar problems
appear in the corresponding dynamical maps. Contrary to our expectations, though, we have not found any difference in 
performance between the weighted and white averaged algorithms. This seems to indicate that even selection of the best performers 
as done by the weighted average is not enough to prevent the algorithm of being trapped in suboptimal solutions, which can only 
be avoided by increasing the number of replica.

It is interesting to note that a variational approach carried out by Kinouchi and
Caticha~\cite{Kinouchi96} was successful in finding the optimal online learning
rule for a perceptron, in the sense that it will saturate the Bayes' generalization bound calculated by
Opper and Haussler~\cite{Opper91}.

Although the perceptron generalization problem is different from the capacity problem, as in the former the dataset is clearly
realizable having been generated by a corresponding perceptron, which might not be the case for the latter; up to the
critical capacity one can assume that the set of random examples, in the typical case, is indeed realizable. In fact, this is
usually one of the underlying assumptions when attempting to solve the capacity problem. This means that we can use the same
algorithms to carry out both tasks.

The precise form for the parallel variational optimal (VO) algorithm for the BIP was derived
in~\cite{Neirotti10} and is given by
\begin{equation}
  \vc{b}(t+1) = \vc{b}(t)+ \frac{\vc{s}_{t}y_t}{\sqrt{N}} F(t),
\end{equation}
where the modulation function is
\begin{equation}
  F(t) = 2 \sqrt{\frac{Q(t)}{R(t)^2}} \col{1-R(t)^2} \frac{\mathcal{N}_t}{1+\erf\prs{R(t)\phi(t)/\sqrt{2(1-R(t)^2)}}},
\end{equation}
with
\begin{equation}
  R(t)    = \frac{\vc{b}_0\cdot\vc{b}(t)}{|\vc{b}_0||\vc{b}(t)|}, \qquad
  Q(t)    = \frac{\vc{b}(t)^2}{N}, \qquad
  \phi(t) = h(t)y_t, \qquad
  h(t)    = \frac{\vc{b}\cdot\vc{s}_t}{|\vc{b}|}~;
\end{equation}
where $\vc{b}_0$ is a teacher perceptron, which in the capacity case would correspond to the correct inferred variable
vector, the true value of which we do not know. In employing the VO algorithm, an assumption that the overlaps are self-averaging 
has been used. Therefore, a sensible way to obtain a value that could
be used as a good estimate of $\vc{b}_0$ is to run the algorithm many times in parallel and average all values of
$\vc{b}(t)$ at each iteration. Like in our algorithm, this average can be either white or weighted.

A notable characteristic of the above set of equations is their similarity with our equations for the OnMP if one substitutes
 \begin{equation}
  m_k\rightarrow b_k, \qquad m_k^2\rightarrow R^2, \qquad R\phi\rightarrow yu, \qquad 1-R^2\rightarrow\sigma^2,
\end{equation}
respectively. In fact, the asymptotic behavior of the VO guarantees that even the square-root amplitude appearing in front of the 
modulation function tends to the same value as in the OnMP, making the two sets isomorphic under this substitution. This striking
relation between both algorithms is a strong indication that our algorithm must also be capable of achieving the optimal capacity 
and saturates Bayes' generalization bound~\cite{Opper91}.

\section{Performance}
\label{section:CP}

In this section we compare the performance of the rOnMP with that of the PT algorithm. The reason for choosing PT is 
that it is a well established parallel algorithm with good performance in searching for solutions in the BIP capacity problem.
Other derivatives of BP-based algorithms have been used to solve the BIP capacity problem, for instance Survey
Propagation~\cite{Braunstein05,Braunstein06}; the latter also aims to address the fragmentation of solution space but employs a
different approach. The results reported~\cite{Braunstein05,Braunstein06} show that solutions can be found very close to the
theoretical limits even for large systems but additional practical techniques and considerations should be used to 
successfully obtain solutions. As our aim in this work is to show how replication can improve significantly the performance of MP
algorithms, we use the PT algorithm as the preferred benchmark method due to its simpler implementation.

Parallel Tempering (PT) or replica exchange Monte Carlo algorithm~\cite{Marinari1992,Swendsen86} was introduced as a tool for
carrying out simulations of spin glasses. Like the BIP capacity problem, spin glasses have a complicated energy landscape with
many peaks and valleys of varying heights and PT has been successfully applied to that and many other similar problems where the
extremely rugged energy landscape causes other methods to underperform \cite{Neirotti00, Neirotti00b}.

In many cases searching for the low energy states is done by gradient descent methods. In statistical physics, simulated
annealing is a principled and useful alternative to gradient descent by allowing for a stochastic search while slowly 
decreasing the temperature; it is particularly effective in the cases where the landscape has one or very few valleys. However, 
to guarantee convergence to an optimal state the temperature should be lowered very slowly and most applications use a much faster 
cooling rate. In the case of spin glasses, this causes the algorithm to be easily trapped in local minima.

The idea behind the PT algorithm is to introduce a number of replica of the system that search the solution space in parallel at
different temperatures using a simple Metropolis-Hastings procedure. The higher the temperature, the easier it is for the replica
to jump over energy barriers, but convergence becomes increasingly compromised. However, jumping over barriers allows for the
exploration of a large part of solution space and the PT algorithm cleverly exploits this by comparing, at chosen time intervals,
the energy of the present random walker at two different successive temperatures. If the higher-temperature random walker reaches
a state of smaller energy than the one at a lower temperature, they are exchanged, otherwise there is an
exponentially small probability for this exchange to take place; this probability is given by the ratio of the Boltzmann 
weights as in the usual Metropolis-Hastings algorithm.

As time proceeds, the lowest energy walker corresponds to the lowest temperature replica. After
convergence or when a certain number of iterations have been done, results for all temperatures can be obtained. PT has
a very high performance for the BIP capacity problem and can achieve very high storage capacities. The disadvantage comes from the fact
that PT uses much more information than is needed to solve the BIP capacity problem and is therefore computationally expensive.

\begin{figure}
  \center
  \includegraphics[width=12cm]{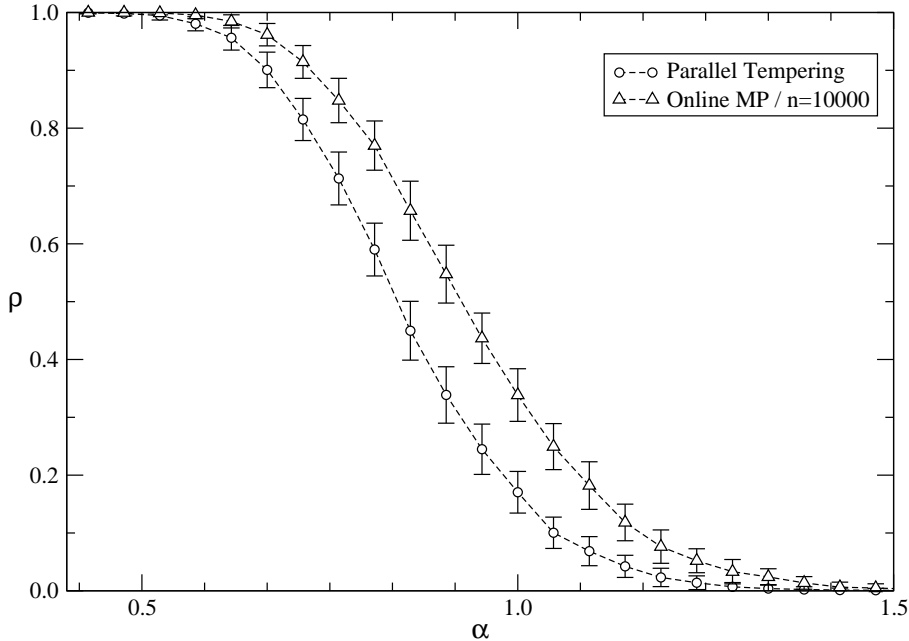}
  \caption{Results from the Parallel Tempering algorithm (circles) versus the replicated online MP (triangles) with 10000 replica.
           The system size is $K=21$. The graph, presented with error bars over 10000 trials, shows the superiority of the
           MP already for this number of replica.}
  \label{figure:PT}
\end{figure}

Figure~\ref{figure:PT} shows the performance of the rOnMP compared to PT for a system size $K=21$.
We see that already with $n=10000$ replica rOnMP has a better performance than PT, which was run up to the point when there was
no extra improvement. We observed that by increasing the number of replica we can reach better performances although the 
improvement in the performance becomes more modest for higher values; studies with a large number of replica $n\sim 10^5$ seem to 
indicate that the critical capacity can indeed be achieved for $n$ sufficiently large. However, the computing time increases as 
well and more lengthy and detailed analysis are necessary to get precise results.Further experiments also seem to indicate that 
in the many replica case the algorithm's performance does not deteriorate with increasing $K$; but clearly, the corresponding 
computing time increases as well.

In addition to the better performance, rOnMP has several other advantages over PT. Firstly, the running time for achieving
a similar performance is lower. Secondly, and more importantly, PT depends on a complicated fine-tuning of the number 
of replica at different temperatures and how these are spaced. Different ranges of temperatures and spacings between them give
different results and these require optimization trials. On the other hand, the application of rOnMP is much more 
straightforward and depends only on the number of replica.

\section{Conclusions}
\label{section:Conclusions}

The main objective for this work was to show that parallelizing message passing algorithms, via \emph{replication} of the
variable system, can lead to a dramatic improvement of their performance. Replication is based on insights and concepts from
statistical physics, especially in the subfield of disordered systems. The binary Ising perceptron (BIP) capacity
problem was chosen as a difficult benchmark problem due to its complex solution space and its discrete output and noise model;
both make the inference problem particularly difficult.

Firstly, we showed analytically that the offline version of the MP algorithm for the BIP capacity problem results in the clipped 
Hebb rule estimator in the thermodynamic limit and when the number of replica is large. This shows a fundamental limitation of 
the MP procedure and motivated us to search for an online version of it; after establishing the single system version it has been 
extended to accommodate a replicated version. Both non-replicated and the replicated versions were shown to have superior 
performance to that of the OffMP.

There are two important aspects of replicated algorithms we would like to point out, namely the way the search is carried out in 
solution space and how to combine the search results to obtain a unified estimate. We devised a method to make replicated variable 
systems follow different paths in the solution space by using different orders of example presentations, which is only possible in 
online algorithms. We also tried two different ways to combine the results, white and weighted averaging, the latter using the 
Boltzmann factors of each replica. We found no difference between both approaches, indicating that weighting the averages is not 
sufficient to avoid local minima.

Finally, we compared the results of the weighted rOnMP algorithm with those of the Parallel Tempering algorithm, showing that our
replicated version of MP performs much better than PT.

Showing that replication in online MP improves its efficiency paves the way to using similar approaches to address other hard
computational problems. We are currently exploring the applicability of techniques developed here to address other problems in 
physics and in information theory. There are still many issues that should be studied concerning these algorithms. One of them, 
which is currently underway, is finding an efficient way to choose the order of examples, which can be seen as a query learning 
procedure. However, query learning for the particular problem studied here corresponds to sampling from a fragmented solution 
space that corresponds to a the replica symmetry breaking solution space and demands the introduction of a carefully constructed 
interaction between the replicated solutions, which we currently investigate.

\section*{Acknowledgments}

Support by the Leverhulme trust (F/00 250/M) is acknowledged.


\bibliographystyle{apsrev4-1}
\bibliography{traj_v02}

\appendix

\section{Message Passing Expansion for the Binary Ising Perceptron}
\label{appendix:MPE}

Consider the first MP equation (\ref{equation:orsp_1}), repeated below for convenience
\begin{equation}
  \label{equation:orsp_1_app}
  \hat{m}^t_{\mu k} = \frac{\sum_{b_k} b_k \mathcal{P}^{t+1}\prs{y_\mu|b_k,\chs{y_{\nu\neq\mu}}}}
                      {\sum_{b_k}\mathcal{P}^{t+1}\prs{y_\mu|b_k,\chs{y_{\nu\neq\mu}}}}.
\end{equation}

We denote the numerator of this expression simply by $A$, ignoring for brevity the dependence on the indices.
By introducing a variable $\xi$ to represent the field $\xi_\mu$ using a Dirac delta, we can write
\begin{equation}
  \begin{split}
    A &= \frac{y_\mu}{2^K} \int\frac{d\xi d\hat{\xi}}{2\pi} \, e^{i\xi\hat{\xi}} (\sgn\xi)
         \col{\prod_{l\neq k} \sum_b (1+m_{\mu l} b) \exp\prs{-i\hat{\xi}\frac{s_{\mu l} b}{\sqrt{K}}}}\\
      &  \quad\times \sum_b b \, \exp\prs{-i\hat{\xi}\frac{s_{\mu k} b}{\sqrt{K}}}.
  \end{split}
\end{equation}

Summing over $b\in\chs{\pm 1}$ one obtains
\begin{equation}
  \begin{split}
    \sum_b (1+m_{\mu k} b) \exp\prs{-i\hat{\xi}\frac{s_{\mu k} b}{\sqrt{K}}}
         &=       2\col{\cos\prs{\frac{\hat{\xi}}{\sqrt{K}}}-im_{\mu k} s_{\mu k} \sin\prs{\frac{\hat{\xi}}{\sqrt{K}}}}\\
         &\approx 2\col{1-im_{\mu k} s_{\mu k}\frac{\hat{\xi}}{\sqrt{K}}-\frac{\hat{\xi}^2}{2K}},
  \end{split}
\end{equation}
where, in the last line, we expand the trigonometric functions to their first non-trivial orders in $1/\sqrt{K}$, already taking 
into consideration the large $K$ scenario. Doing the same expansion to the second summation one obtains
\begin{equation}
  \begin{split}
    \sum_b b \, \exp\prs{-i\hat{\xi}\frac{s_{\mu k} b_k}{\sqrt{K}}} &=       -2i s_{\mu k} \sin\prs{\frac{\hat{\xi}}{\sqrt{K}}}\\
                                                                    &\approx -2i s_{\mu k}\frac{\hat{\xi}}{\sqrt{K}},
  \end{split}
\end{equation}

These approximations allow one to rewrite the expression for $A$ as
\begin{equation}
  \begin{split}
    A &=       \frac{-iy_\mu s_{\mu k}}{\sqrt{K}} \int\frac{d\xi d\hat{\xi}}{2\pi} \, e^{i\xi\hat{\xi}} (\sgn\xi) \hat{\xi}
               \exp\col{\sum_l \ln \prs{1-\frac{\hat{\xi}^2}{2K}-im_{\mu l} s_{\mu l}\frac{\hat{\xi}}{\sqrt{K}}}}\\
      &\approx \frac{-iy_\mu s_{\mu k}}{\sqrt{K}} \int\frac{d\xi}{2\pi} \, (\sgn\xi) \int d\hat{\xi}\,\hat{\xi}
                         \exp\col{-\frac{\hat{\xi}^2\sigma^2_{\mu k}}{2}+i\hat{\xi}\prs{\xi-u_{\mu k}}},
  \end{split}
\end{equation}
where
\begin{equation}
  \sigma^2_{\mu k} = \frac1K \sum_{l\neq k} (1-m_{\mu l}^2), \qquad 
                     u_{\mu k} = \frac1{\sqrt{K}} \sum_{l\neq k} m_{\mu l} s_{\mu l}.
\end{equation}

The resulting integral is trivial and, by following the analogous steps for the denominator, we finally reach the result given by
expression (\ref{equation:MPFE}).

\section{Analytical Derivation of the Replicated Naive MP Algorithm}
\label{appendix:Hebb}

Upon replication of the variable system such that the final estimate of the variable vector is inferred by a white average of 
the $n$ replica
\begin{equation}
  \hat{b}_k = \sgn\prs{\frac1n \sum_{a=1}^n b_k^a},
\end{equation}
one can take the limit $n\rightarrow\infty$ to calculate a closed expression for it. The MP equations (\ref{equation:orsp_1}) and
(\ref{equation:orsp_2}) remain the same, but the likelihood term has to include the contribution of the replica as
\begin{equation}
  \prob{y_\mu|\vc{b}} = \sum_{\chs{\vc{b}^a}} \prob{y_\mu|\vc{b},\chs{\vc{b}^a}}\prob{\chs{\vc{b}^a}|\vc{b}},
\end{equation}
\begin{align}
  \prob{y_\mu|\vc{b}}          &=     \frac1{2^{n+1}} \col{1+y_\mu\,\sgn\prs{\frac1{\sqrt{K}} \sum_{k=1}^K s_{\mu k} b_k}}
                                      \prod_a\col{1+y_\mu\,\sgn\prs{\frac1{\sqrt{K}} \sum_{k=1}^K s_{\mu k} b^a_k}},\\
  \prob{\chs{\vc{b}^a}|\vc{b}} &\propto \prod_{k} \frac12 \col{1+b_k\,\sgn\prs{\frac1n \sum_{a=1}^n b_k^a}}.
\end{align}

In the last equation we ignore the normalization. For the calculation to be carried out rigorously, the normalization should be 
taken into account in what follows. However, careful calculations show that it does not change the saddle point result. The
above expressions can be substituted in the first of the MP equations (\ref{equation:orsp_1}). Let us concentrate on the numerator 
of Eq.~(\ref{equation:orsp_1}), which can be written as
\begin{equation}
  \begin{split}
    A &\propto \int \col{\frac{d\xi d\hat{\xi}}{2\pi} e^{i\xi\hat{\xi}}}
               \col{\prod_a \frac{d\xi^a d\hat{\xi}^a}{2\pi} e^{i\xi^a\hat{\xi}^a}} \prs{1+y_\mu\,\sgn\xi}
               \prod_a \prs{1+y_\mu\,\sgn\xi^a}\\
      &        \quad\times\sum_{\vc{b}} b_k \col{\prod_{l\neq k}\frac12 \prs{1+b_l m_{\mu l}}}
               \exp\col{-\frac{i\hat{\xi}}{\sqrt{K}}\sum_{j=1}^K s_{\mu j} b_j}\\
      &        \quad\times\sum_{\chs{\vc{b}^a}} \prod_j \frac12 \col{1+b_j \,\sgn\prs{\frac1n \sum_{a=1}^n b_j^a}}
               \exp\col{-\frac{i}{\sqrt{K}}\sum_a \hat{\xi}^a \sum_{j=1}^K s_{\mu j} b_j^a}.
  \end{split}
\end{equation}

To decouple the replicated systems, we introduce the $K$ variables
\begin{equation}
  \lambda_k = \frac1n \sum_a b_k^a,
\end{equation}
via Dirac deltas. By defining the notation
\begin{equation}
  D\col{\xi,\hat{\xi}} \equiv \col{\frac{d\xi d\hat{\xi}}{2\pi} e^{i\xi\hat{\xi}}}
                              \col{\prod_a \frac{d\xi^a d\hat{\xi}^a}{2\pi} e^{i\xi^a\hat{\xi}^a}}, \qquad
  D\col{\lambda,\hat{\lambda}} \equiv \col{\prod_k \frac{d\lambda_k d\hat{\lambda}_k}{2\pi/n} e^{in\lambda_k\hat{\lambda}_k}},
\end{equation}
and summing over $\vc{b}$'s we obtain
\begin{equation}
  \begin{split}
    A &\propto \int D\col{\lambda,\hat{\lambda}} D\col{\xi,\hat{\xi}} \prs{1+y_\mu\,\sgn\xi}
               \prod_a \prs{1+y_\mu\,\sgn\xi^a}\\
      &        \quad\times\col{\prod_{a,j} \cos\prs{\hat{\lambda}_j+\frac{\hat{\xi}^a s_{\mu j}}{\sqrt{K}}}}
               \col{-i\sin\prs{\frac{\hat{\xi} s_{\mu k}}{\sqrt{K}}}+\sgn\lambda_k\cos\prs{\frac{\hat{\xi} s_{\mu k}}{\sqrt{K}}}}\\
      &        \quad\times\prod_{l\neq k} \prs{1+m_{\mu l}\,\sgn \lambda_l}\col{ \cos\prs{\frac{\hat{\xi}s_{\mu l}}{\sqrt{K}}}
               -i\,\sgn\lambda_l\sin\prs{\frac{\hat{\xi}s_{\mu l}}{\sqrt{K}}}}.
  \end{split}
\end{equation}

One can now expand the arguments of the $\cos$ and $\sin$ functions in powers of $1/\sqrt{K}$ to obtain
\begin{equation}
  \begin{split}
    A &\propto \int D\col{\lambda,\hat{\lambda}} D\col{\xi,\hat{\xi}} \prs{1+y_\mu\,\sgn\xi}
               \prod_a \prs{1+y_\mu\,\sgn\xi^a}\\
      &        \quad\times \exp\col{\sum_{a,j} \ln \prs{\cos\hat{\lambda}_j
               -\frac{\hat{\xi}^a s_{\mu j}}{\sqrt{K}}\sin\hat{\lambda}_j -\frac{(\hat{\xi}^a)^2}{2K} \cos\hat{\lambda}_j}}
               \prs{-i\frac{\hat{\xi}s_{\mu k}}{\sqrt{K}}+\sgn\lambda_k}\\
      &        \quad\times\exp\col{\sum_{l\neq k}\ln\prs{1+m_{\mu l}\,\sgn \lambda_l}
               +\sum_{l\neq k} \ln\prs{1-\frac{\hat{\xi}^2}{2K}-\frac{i\hat{\xi}}{\sqrt{K}}\,\sgn\lambda_l}}.
  \end{split}
\end{equation}

The integrals over the $\xi$ variables are easy to calculate, leading to the following expression at leading order in $1/\sqrt{K}$
\begin{equation}
\label{eq:saddlepointintegral}
  A \propto \int \col{\prod_j \frac{d\lambda_j d\hat{\lambda}_j}{2\pi/n}} \sgn\lambda_k \, e^{n\Phi},
\end{equation}
where
\begin{equation}
  \Phi = i\sum_j \lambda_j\hat{\lambda}_j +\frac1n \sum_{l\neq k} \ln\prs{1+m_{\mu l} \sgn \lambda_l} +\sum_j \ln \cos\hat{\lambda}_j
         +\frac1n\sum_{c=0}^n \ln I_c,
\end{equation}
with
\begin{align}
  I_a          &= 1+y_\mu \erf\prs{\frac{u_{\mu}}{\sqrt{2\sigma_\mu^2}}}, \qquad a=1,...,n\\
  u_\mu        &= -\frac{i}{\sqrt{K}}\sum_j s_{\mu j} \tan \hat{\lambda}_j,\\
  \sigma^2_\mu &= \frac1K \sum_j \prs{1+\tan^2 \hat{\lambda}_j},\\
  I_0          &= 1+y_\mu\erf\prs{\frac{u^0_{\mu k}}{\sqrt{2}}},\\
  u^0_{\mu k}  &= \frac1{\sqrt{K}} \sum_{l\neq k} s_{\mu l}\,\sgn\lambda_l.
\end{align}

Following the same calculations for the denominator, one can see that for large $n$ the variables $\hat{m}_{\mu k}$ are given by
$\sgn \lambda^*_k$, where $\lambda^*_k$ is defined by the saddle point of the integral (\ref{eq:saddlepointintegral}) which is a 
solution of the simultaneous equations
\begin{equation}
  \pder{\Phi}{\lambda_j}=\pder{\Phi}{\hat{\lambda}_j}=0.
\end{equation}

Differentiating $\Phi$ we finally find the result
\begin{equation}
  \hat{m}_{\mu k}=y_\mu s_{\mu k},
\end{equation}
resulting in the estimate (\ref{equation:Hebb}) of the variable vectors that also corresponds to the clipped Hebb rule.

\end{document}